\documentclass[journal,11pt,draftclsnofoot,onecolumn]{IEEEtran}
\usepackage{cite}

\ifCLASSINFOpdf
\else
\fi
\usepackage{amsmath}
\usepackage{array}
\usepackage{url}

\renewcommand{\IEEEQED}{\IEEEQEDopen}

\usepackage{amssymb}
\usepackage{color}
\newtheorem{definition}{Definition}[section]
\newtheorem{theorem}{Theorem}[section]
\newtheorem{corollary}{Corollary}[section]
\newtheorem{lemma}{Lemma}[section]

\newtheorem{remark}{Remark}[section]

\begin{document}
%
\title{Optimum Self Random Number Generation Rate and Its Application to Rate Distortion Perception Function}
%
%
%

\author{Ryo Nomura,~\IEEEmembership{Member,~IEEE}%
\thanks{R. Nomura is with the Center for Data Science, Waseda University, Tokyo 169-8050, Japan, e-mail: nomu@waseda.jp}
\thanks{This paper is an extended version of the conference paper \cite{Nomura_ISIT2023}.
Compared to [1], it includes the complete proof of the theorem in Chapter 4, adds Theorem 5.2, which is one of the main results, and also incorporates Chapter 6 and three appendices.
}
}

\maketitle

\begin{abstract}
The self-random number generation (SRNG) problem is considered for \textit{general} setting.
In the literature, the optimum SRNG rate with respect to the variational distance has been discussed. 
In this paper, we first try to characterize the optimum SRNG rate with respect to a subclass of $f$-divergences.
The subclass of $f$-divergences considered in this paper includes typical distance measures such as the variational distance, the KL divergence, the Hellinger distance and so on. Hence our result can be considered as a generalization of the previous result with respect to the variational distance.
Next, we consider the obtained optimum SRNG rate from several viewpoints.
The $\varepsilon$-source coding problem is one of related problems with the SRNG problem.
Our results reveal how the SRNG problem with the $f$-divergence relate to the $\varepsilon$-fixed-length source coding problem.
We also apply our results to the rate distortion perception (RDP) function.
As a result, we can establish a lower bound for the RDP function with respect to $f$-divergences using our findings. 
Finally, we discuss the representation of the optimum SRNG rate using the smooth R\'enyi entropy.
\end{abstract}


%
\IEEEpeerreviewmaketitle

\section{Introduction}
%
%
%
%
\IEEEPARstart{I}n information theory, the random number generation problem focuses on approximating a given probability distribution (target distribution), using another probability distribution (coin distribution) \cite{VV,HH97,VKV98,Han2005,Hayashi,NH2011,YT2019,Nomura_ITW2020,WH2020}.
In this study, we examine the scenario wherein the target distribution is identical to the coin distribution.
This problem is called the \textit{self-random number generation} (SRNG) problem \cite{Han}.
The primary goal of the SRNG problem is to efficiently approximate the source ${\bf X} = \{ X^n \}_{n=1}^\infty$ using the source itself, ensuring that the approximation error remains smaller than or equal to a specified constant.
To formulate this problem, we introduce an approximation (distance) measure, denoted by $d(X^n,Y^n)$ to quantify the distance between two probability distributions $P_{X^n}$ and $P_{Y^n}$.
Given an arbitrary general source ${\bf X} = \{ X^n \}_{n=1}^\infty$, our aim is to find a mapping $\phi_n(X^n)$ satisfying $\limsup_{n \to \infty}d(X^n,\phi_n(X^n)) \le D$. The objective here is to minimize the rate $\log |\phi_n|$ adhering to the aforementioned condition. 
The determination of this optimum achievable rate is our primary focus.

The investigation of the optimum achievable rate in the SRNG problem, particularly in relation to the variational distance, has been conducted by Han \cite[Sect. 2.6]{Han}. The proof indicates that all fixed-length source codes with diminishing error probabilities qualify as self-random number generators for the source, in terms of variational distance.
Consequently, this perspective reveals a close relationship between the SRNG problem and the source coding problem.
A similar problem setting has also been addressed by Kumagai and Hayashi \cite{KH2013,Kumagai2017a,Kumagai2017b} in which a distance measure related to the Hellinger distance has been utilized.
Especially \cite{Kumagai2017a,Kumagai2017b} have examined it within a broader framework of general random number generation problem.
Their main discussion centers around the second-order asymptotics for i.i.d. (independent and identically distributed) sources.

However, in the SRNG problem the optimum achievable rate with respect to other approximation measures has not been considered yet while many approximation measures exist such as Kullback-Leibler (KL) divergence, Hellinger distance, and so on.
To address the optimum SRNG rate in relation to these alternative approximation measures, our study delves into a class of $f$-divergences.
The $f$-divergence is a general distance measure \cite{csiszar2004information,SV2016}, which includes several important measures such as the variational distance, the KL divergence, the Hellinger distance, and so on.
Therefore, exploring the SRNG problem within the context of the $f$-divergence is a significant and relevant endeavor.


Next, we apply our general formula regarding the optimum achievable rate into the rate distortion perception (RDP) problem.
Recently, the RDP problem has been paid attention in the lossy source coding framework \cite{BM2019,LW2021,JC2022}.
The notion of perception quality has been proposed by Blau and Michaeli \cite{BM2019} in the image processing area. 
They have defined perception quality as the probabilistic distance between the original information source and the reconstructed information.  
We consider the $f$-divergence as the perception quality and derive the lower bound of the RDP function.

We also consider another expression of the optimum achievable rate in the SRNG problem using the smooth R\'enyi entropy.
The smooth R\'enyi entropy of the source is an information quantity which is often used to address coding theorems in information theory \cite{RW04,Uyematsu2010}.
This is beneficial as it provides a perspective that is distinct from the traditional representation.
Hence, in this paper, we try to express the optimum achievable rate using the smooth R\'enyi entropy of the source.

This paper is organized as follows. 
In Section II, we define the SRNG problem with respect to $f$-divergences and define the achievable rate. Then, we review the previous result on the optimum achievable rate. In Section III, we derive two fundamental finite blocklength lemmas that reveal relationship between the $f$-divergences and the rate of mappings.
In Section IV, based on two lemmas established in Section III, we derive the optimum achievable rate in the SRNG problem. Then, we compute the optimum achievable rates for some specified distance measures.
In Section V, an application to the RDP function is considered to elucidate the significance of Theorem 4.1. In Section VI, we establish the formula for the optimum achievable rate using the smooth max entropy. Finally, we conclude our results in Section VII.
%
%
%
%
\section{Preliminaries}
We define the \textit{general} source as an infinite sequence 
\begin{equation}
{\bf X} = \left\{X^n = \left(X_1^{(n)}, X_2^{(n)}, \ldots, X_n^{(n)}\right) \right\}_{n=1}^\infty
\end{equation}
of $n$-dimensional random variables $X^n$. Here, each component random variable $X^{(n)}_i$ takes values in a finite or infinite countable set ${\cal X}$ \cite{Han}. 
Our discussion centers on how to efficiently approximate the source ${\bf X}$ using ${\bf X}$ itself.
Let us consider two mappings $\varphi_n:{\cal X}^n \to {\cal M}_n := \{1,2,\dots,M_n \}$ and $\psi_n:{\cal M}_n \to {\cal X}^n$, and set $\tilde{X}^n=\psi_n(\varphi_n({X^n}))$.
First, we consider the variational distance
\begin{equation}  \label{eq:vd}
d(X^n||\tilde{X}^n) := \frac{1}{2} \sum_{{\bf x} \in {\cal X}^n } |P_{X^n}({\bf x}) - P_{\tilde{X}^n}({\bf x})  |
\end{equation}
as an approximation measure.
\begin{definition}
Rate $R$ is said to be achievable with the variational distance if there exists a sequence of mappings $(\varphi_n,\psi_n)$ such that 
\begin{equation}
\lim_{n \rightarrow \infty} d(X^n||\tilde{X}^n) =0 \mbox{ and } \limsup_{n \rightarrow \infty}  \frac{1}{n} \log M_n \leq R.
\end{equation}
\end{definition}
\begin{definition}[Optimum SRNG rate with VD]
\begin{equation*}
S_0({\bf X}) =  \inf \left\{ R | \mbox{$R$ is achievable with the variational distance} \right\}.
\end{equation*}
\end{definition}
The following theorem has been shown by Han \cite{Han}.
\begin{theorem}[Han \cite{Han}] \label{theo:1-3}
\begin{equation}
S_0({\bf X}) = \overline{H}({\bf X}),
\end{equation}
where $\overline{H}({\bf X})$ is called the {\it spectral sup-entropy rate} of the source ${\bf X}$ \cite{Han} defined as
\begin{equation}
 \overline{H}({\bf X}) := \inf \left\{  R \left| \lim_{n \to \infty } \Pr \left\{ \frac{1}{n} \log \frac{1}{P_{X^n}(X^n)}  \!>\!  R  \right\} = 0 \right. \right\}.
\end{equation}
\end{theorem}
The proof of this theorem reveals that all fixed-length source codes exhibiting diminishing error probabilities can be regarded as self-random number generators for the source.
Consequently, this result elucidates a specific type of relationship between the source coding problem and the random number generation problem.

We aim to extend these findings to include the case of $f$-divergences.
The $f$-divergence between two probabilistic distributions $P_{Z}$ and $P_{\overline{Z}}$ is defined as follows \cite{csiszar2004information}.
Let $f(t)$ be a convex function defined for $t >0$ and $f(1) =0$.
\begin{definition}
Let $P_Z$ and $P_{\overline{Z}}$ denote probability distributions over a finite or countably infinite set ${\cal Z}$. The $f$-divergence between $P_{Z}$ and $P_{\overline{Z}}$ is defined by
\begin{equation}
D_f(Z||\overline{Z}) :=  \sum_{z \in {{\cal Z}}} P_{\overline{Z}}(z) f \left(\frac{P_{{Z}}(z)}{P_{\overline{Z}}(z)}\right),
\end{equation}
where we set $0 f\left(\frac{0}{0}\right) =0$, $f(0) = \lim_{t \to 0} f(t)$, $0f(\frac{a}{0}) = \lim_{t \to 0} tf(\frac{a}{t}) = a \lim_{u \to \infty}\frac{f(u)}{u}$.
\end{definition}
We give some examples of $f$-divergences \cite{csiszar2004information,SV2016}:
\begin{itemize}
\item $f(t) = t \log t$:  (Kullback-Leibler divergence)
\begin{IEEEeqnarray}{rCl}
D_f(Z||\overline{Z}) =   \sum_{z \in {{\cal Z}}} P_{{Z}}(z) \log \frac{P_{{Z}}(z)}{P_{\overline{Z}}(z)} =: D(Z||\overline{Z}).
\end{IEEEeqnarray}
\item $f(t) = - \log t$:  (Reverse Kullback-Leibler divergence)
\begin{IEEEeqnarray}{rCl}
D_f(Z||\overline{Z}) =   \sum_{z \in {{\cal Z}}} P_{\overline{Z}}(z) \log \frac{P_{\overline{Z}}(z)}{P_{{Z}}(z)} = D(\overline{Z}||{Z}).
\end{IEEEeqnarray}
\item $f(t) = 1 - \sqrt{t}$:  (Hellinger distance)
\begin{IEEEeqnarray}{rCl}
D_f(Z||\overline{Z}) =   1 - \sum_{z \in {{\cal Z}}}\sqrt{P_{{Z}}(z) P_{\overline{Z}}(z)}.
\end{IEEEeqnarray}
\item $f(t) = (1-t)^+ := \max\{1-t,0\}$:  (Variational distance)
\begin{IEEEeqnarray}{rCl}  \label{eq:hvd}
D_f(Z||\overline{Z}) =  \frac{1}{2} \sum_{z \in {{\cal Z}}}  |(P_{{Z}}(z) - P_{\overline{Z}}(z))|.
\end{IEEEeqnarray}
\item $f(t) = (t-\gamma)^+$ :  ($E_\gamma$-divergence)
For any given $\gamma \ge1$, 
\begin{IEEEeqnarray}{rCl} \label{eq:egamma}
D_f(Z||\overline{Z}) =  \sum_{z \in {{\cal Z}}: P_{{Z}}(z) > \gamma P_{\overline{Z}}(z)} \left(P_{{Z}}(z) - \gamma P_{\overline{Z}}(z)\right). 
\end{IEEEeqnarray}
\end{itemize}
\begin{remark}
The $E_\gamma$-divergence between two probabilistic distribution $P_Z$ and $P_{\overline{Z}}$ is defined as (\ref{eq:egamma}) \cite{SV2016,LCV2017}.
However, from the simple observation tells us the alternative expression of $E_{\gamma}$-divergence using the function \cite{Nomura_TIT2020}:
\begin{equation} \label{eq:egamma2}
f(t) = (\gamma - t)^+ + 1 -\gamma.
\end{equation}
\end{remark}

From Jensen's inequality the following key property holds for $f$-divergences. Let $a: {\cal Z} \to \mathbb{R}^+$ and $b : {\cal Z} \to \mathbb{R}^+$ denote two functions. Then, it holds that
\begin{IEEEeqnarray}{rCl} \label{eq:log-sum}
\sum_{z \in {\cal Z}'} b(z) f\left(\frac{a(z)}{b(z)} \right) \! \ge\! \left(\sum_{z \in {\cal Z}'} b(z) \right)  f\left(\frac{\sum_{z \in {\cal Z}'}a(z)}{\sum_{z \in {\cal Z}'} b(z)} \right).
\end{IEEEeqnarray}
In this study, we assume the following condition on the function $f$.
\begin{description}
\item[C1)] The function $f(t)$ is a monotonically decreasing function of $t$. That is, for any pair of positive real numbers $(a,b)$ satisfying $a < b$ it holds that
$f(a) \ge f(b)$.
\item[C2)] For any pair of positive real numbers  $(a,b)$, it holds that
\begin{equation} \label{assumption:3}
\lim_{n\to \infty} \frac{f\left(e^{-nb} \right)}{e^{na}} = 0.
\end{equation}
\item[C3)] For any positive number $a \in [0,1]$, it holds that
\begin{equation} \label{assumption:4}
0 f\left(\frac{a}{0}\right) = 0.
\end{equation}
\end{description}
\begin{remark}  \label{remark:conditions}
Notice here that $f(t)= - \log t$, $f(t)= 1- \sqrt{t}$ and $f(t)= ( 1- t)^+$ are satisfy these three conditions, while $f(t)= t\log t$ does not.
Moreover, it is not difficult to check that (\ref{eq:egamma2}) satisfies these conditions.
\end{remark}
\begin{remark}  \label{remark:conditions2}
If we consider the case where $\max_{t \in (0,+\infty]}f(t) < + \infty$ holds, then C2) holds. 
Hence, the condition C2) controls the convergence rate of $f(t)$ as $t$ approaches $0$.
\end{remark}

\section{Fundamental lemmas}
Before considering optimum achievable rates in the SRNG problem with $f$-divergences, we show two useful lemmas which reveal some kind of relationships between $f$-divergences and a pair of mappings $(\varphi_n:{\cal X}^n \to {\cal M}_n$, $\psi_n:{\cal M}_n \to {\cal X}^n)$.
\begin{lemma} \label{lemma:direct}
Assuming that the function $f$ satisfies conditions C1) and C3), for any $M_n$ and $\gamma >0$, there exists a pair of mappings $(\varphi_n,\psi_n)$
satisfying
\begin{IEEEeqnarray}{rCl}
\lefteqn{D_f(X^n||\psi_n(\varphi_n({X^n})))} \nonumber \\
&  \le  & 
f\left( \Pr \left\{ \frac{1}{n} \log \frac{1}{P_{X^n}(X^n)}  \le  \frac{1}{n}\log M_n  - \gamma  \right\}  -   e^{- n\gamma} \right) + e^{-n\gamma}f\left(\frac{1}{M_n} \right).
\end{IEEEeqnarray}
\end{lemma}
\begin{IEEEproof} 
We define two sets $S_n$ and $T_n$ as follows
\begin{IEEEeqnarray*}{rCl} 
S_n := \left\{ {\bf x} \in {\cal X}^n \left| \frac{1}{n} \log \frac{1}{P_{X^n}({\bf x})} \le \frac{1}{n}\log M_n - {\gamma}\right. \right\},
\end{IEEEeqnarray*}
\begin{IEEEeqnarray*}{rCl} 
T_n \! := \! \left\{ {\bf x} \! \in \! {\cal X}^n \left| \frac{1}{n}\log M_n \! - \! {\gamma} \! < \! \frac{1}{n} \log \frac{1}{P_{X^n}({\bf x})} \! \le \! \frac{1}{n}\log M_n\right. \right\}.
\end{IEEEeqnarray*}
Arrange elements in $S_n \cup T_n $ according to $P_{X^n}$ in descending order and set
\begin{equation}
\{{\bf x}_1,{\bf x}_2,\dots, {\bf x}_{|S_n|},{\bf x}_{|S_n|+1},\dots, {\bf x}_{|S_n|+|T_n|}  \}
.
\end{equation}
Let a complement set $U_n$ as $U_n = (S_n \cup T_n)^c$. Then, for any ${\bf x} \in U_n$ we have
\begin{equation}  \label{eq:a-13}
P_{X^n}({\bf x})  <  \frac{1}{M_n}.
\end{equation}
Furthermore, since
\begin{equation}
1 \ge \sum_{{\bf x} \in S_n} P_{X^n}({\bf x})  \ge  \left|S_n\right|  \frac{1}{M_n} e^{n\gamma}
\end{equation}
holds, we have
\begin{IEEEeqnarray}{rCl} \label{eq:a-9-1}
\left|S_n \right| \le M_n e^{-n\gamma}.
\end{IEEEeqnarray}
Similarly, we have $\left|S_n \cup T_n \right| \le M_n$.

Here, let $P_{\overline{X}^n}$ denote the probability distribution over $S_n$ defined by
\begin{IEEEeqnarray}{rCl} \label{eq:1-0}
P_{\overline{X}^n}({\bf x}) = \left\{ \begin{array}{ll}
\frac{P_{X^n}({\bf x})}{\Pr\left\{ X^n \in S_n \right\}} & {\bf x} \in S_n, \\
{0} & \mbox{otherwise} .
\end{array} \right.
\end{IEEEeqnarray}
In order to construct mappings, we choose a set $A_n(i) \subseteq U_n$ for ${\bf x}_i \in S_n$ as follows.
For ${\bf x}_1 \in S_n$ we choose arbitrary set of sequences $A_n(1) \subseteq U_n$ such that
\begin{equation}
P_{{X}^n}({\bf x}_1) + \sum_{{\bf x} \in A_n(1)}P_{{X}^n}({\bf x})  \le P_{\overline{X}^n}({\bf x}_1)
\end{equation}
and 
\begin{equation}
P_{{X}^n}({\bf x}_1) + \sum_{{\bf x} \in A_n(1)}P_{{X}^n}({\bf x}) + P_{{X}^n}({\bf x}')  > P_{\overline{X}^n}({\bf x}_1)
\end{equation}
for any ${\bf x}' \in U_n \setminus A_n(1)$.

Secondly, for $ {\bf x}_2 \in S_n$ we choose arbitrary set of sequences $A_n(2) \subseteq U_n\setminus A_n(1)$ such that
\begin{equation}
P_{{X}^n}({\bf x}_2) + \sum_{{\bf x} \in A_n(2)}P_{{X}^n}({\bf x})  \le P_{\overline{X}^n}({\bf x}_2)
\end{equation}
and 
\begin{equation}
P_{{X}^n}({\bf x}_2) + \sum_{{\bf x} \in A_n(2)}P_{{X}^n}({\bf x}) + P_{{X}^n}({\bf x}')  > P_{\overline{X}^n}({\bf x}_2)
\end{equation}
for any ${\bf x}' \in U_n \setminus (A_n(1) \cup A_n(2) )$.
In the similar way, we repeat this operation to choose $A_n(i)$ for ${\bf x}_i$ as long as possible. 
Suppose that this operation stops at $i_0$. 
We set $A_n(i_0) =  U_n \setminus \bigcup_{i=1}^{i_0-1} A_n(i)$.
Here, noting that $\Pr\{ X^n \in (S_n \cup U_n) \} \le 1$ holds,  $i_0$ is smaller than or equal to $|S_n|$ from the construction.
If $i_0 < |S_n|$ holds, we set $A_n(j) = \emptyset$ for all $i_0 < j \le |S_n|$. 
For a sequence ${\bf x}_{|S_n|+j} \ (1 \le \forall j \le |T_n|) $ we set $A_n({|S_n|+j}) = \emptyset$ as well.


Now, we define two mappings $\varphi_n: {\cal X}^n \to {\cal M}_n$ and $\psi_n: {\cal M}_n \to {\cal X}^n$ by using $A_n(i)$ as follows
\begin{IEEEeqnarray}{rCl}
\varphi_n({\bf x}) = i, \mbox{ for }  {\bf x} \in \left( \{{\bf x}_i\} \cup A_n(i) \right),
\end{IEEEeqnarray}
and
\begin{equation}
\psi_n(i)  =  {\bf x}_i.
\end{equation}

Next, we evaluate the performance of these mappings. 
We use the notation $\tilde{X}^n = \psi_n(\varphi_n({X^n}))$ for short.

From the construction of the mapping and (\ref{eq:a-13}), for all $i$ satisfying $1 \le  i\le i_0-1$ it holds that 
\begin{equation} \label{eq:a-11}
P_{\tilde{X}^n} ({\bf x}_i) \le P_{\overline{X}^n} ({\bf x}_i)  < P_{\tilde{X}^n} ({\bf x}_i) + \frac{1}{M_n}.
\end{equation}
On the other hand, from the construction of the mapping we obtain
\begin{IEEEeqnarray}{rCl}  \label{eq:a-3}
\lefteqn{P_{\tilde{X}^n} ({\bf x}_{i_0}) - P_{\overline{X}^n} ({\bf x}_{i_0}) } \nonumber \\
& = & 1\! - \! \Pr\{X^n \in T_n\} \!-\! \sum_{i=1}^{i_0-1}P_{\tilde{X}^n} ({\bf x}_{i}) - \left(1\! - \!\sum_{i=1}^{i_0-1}P_{\overline{X}^n} ({\bf x}_{i})\right) \nonumber \\
& \le & \sum_{i=1}^{i_0-1} \left( P_{\overline{X}^n} ({\bf x}_{i}) - P_{\tilde{X}^n} ({\bf x}_{i}) \right) \nonumber \\
& < & \frac{|S_n|}{M_n} \le e^{-n\gamma},
\end{IEEEeqnarray}
where the last inequality is due to (\ref{eq:a-9-1}).

Furthermore, it is not difficult to check that it holds that 
\begin{equation} \label{eq:a-3-2}
P_{\tilde{X}^n} ({\bf x}_{i}) = P_{{X}^n} ({\bf x}_{i})
\end{equation}
for all $i$ satisfying $i_0+1 \le  i \le |S_n|+|T_n|$, and 
$
P_{\tilde{X}^n} ({\bf x}_{i}) = 0
$
for all $i > |S_n|+|T_n|$.

Thus, noting that the condition C3) and $f(1)=0$, the $f$-divergence between $P_{\tilde{X}^n}$ and $P_{X^n}$ is evaluated as follows:
\begin{IEEEeqnarray}{rCl}  \label{eq:a-4}
D_f\left( {X^n}||\tilde{X}^n \right)
 & = & \sum_{i=1}^{|S_n|+|T_n|} P_{\tilde{X}^n} ({\bf x}_i) f \left(\frac{P_{{X}^n} ({\bf x}_i)}{P_{\tilde{X}^n} ({\bf x}_i)} \right) \nonumber \\
& = & \sum_{i=1}^{i_0-1} P_{\tilde{X}^n} ({\bf x}_i) f\left( \frac{P_{{X}^n} ({\bf x}_i)}{P_{\tilde{X}^n} ({\bf x}_i) } \right) + P_{\tilde{X}^n} ({\bf x}_{i_0}) f\left( \frac{P_{{X}^n} ({\bf x}_{i_0})}{P_{\tilde{X}^n} ({\bf x}_{i_0}) } \right) \nonumber \\
& = & \sum_{i=1}^{i_0-1} P_{\tilde{X}^n} ({\bf x}_i) f\left( \frac{P_{\overline{X}^n} ({\bf x}_i)\Pr\{X^n \in S_n\}}{P_{\tilde{X}^n} ({\bf x}_i) } \right)  \nonumber \\
& & + P_{\tilde{X}^n} ({\bf x}_{i_0}) f\left( \frac{P_{{X}^n} ({\bf x}_{i_0})}{P_{\tilde{X}^n} ({\bf x}_{i_0}) } \right) \nonumber \\
& \le & \sum_{i=1}^{i_0-1} P_{\tilde{X}^n} ({\bf x}_i) f\left( \Pr\{X^n \in S_n\} \right) \nonumber \\
&&+ \left( P_{\overline{X}^n} ({\bf x}_{i_0}) \!+ \! e^{-n\gamma} \right) f\left( \frac{P_{\overline{X}^n} ({\bf x}_{i_0})\Pr\{X^n \in S_n\}}{P_{\overline{X}^n} ({\bf x}_{i_0}) + e^{-n\gamma}} \right),
\end{IEEEeqnarray}
where the second and the third equality is due to (\ref{eq:a-3-2}) and (\ref{eq:1-0}), respectively, and the last inequality is due to (\ref{eq:a-3}).

Next, we evaluate the second term on the RHS of the above inequality.
Using a relation
\begin{IEEEeqnarray}{rCl}
P_{\overline{X}^n} ({\bf x}_{i_0})
& = & ( 1\! - \! e^{-n\gamma})P_{\overline{X}^n} ({\bf x}_{i_0}) +e^{-n\gamma} P_{\overline{X}^n} ({\bf x}_{i_0}),
\end{IEEEeqnarray}
from (\ref{eq:log-sum}) we have
\begin{IEEEeqnarray}{rCl}  \label{eq:a-5}
\lefteqn{\left( P_{\overline{X}^n} ({\bf x}_{i_0}) + e^{-n\gamma} \right) f\left( \frac{P_{\overline{X}^n} ({\bf x}_{i_0})\Pr\{X^n \in S_n\}}{P_{\overline{X}^n} ({\bf x}_{i_0}) + e^{-n\gamma}} \right) } \nonumber \\
& \le & P_{\overline{X}^n} ({\bf x}_{i_0}) f\left( \frac{(1- e^{-n\gamma})  P_{\overline{X}^n} ({\bf x}_{i_0})\Pr\{X^n \in S_n\}}{P_{\overline{X}^n} ({\bf x}_{i_0})} \right)  + e^{-n\gamma} f\left( \frac{e^{-n\gamma} P_{\overline{X}^n} ({\bf x}_{i_0})\Pr\{X^n \in S_n\}}{ e^{-n\gamma}} \right) \nonumber \\
& = &  P_{\overline{X}^n} ({\bf x}_{i_0}) f\left( (1- e^{-n\gamma}) \Pr\{X^n \in S_n\} \right) + e^{-n\gamma} f\left( P_{\overline{X}^n} ({\bf x}_{i_0})\Pr\{X^n \in S_n\} \right) \nonumber \\
& \le & P_{\overline{X}^n} ({\bf x}_{i_0}) f\left(  \Pr\{X^n \in S_n\} -  e^{-n\gamma} \right) + e^{-n\gamma} f\left( \frac{1}{M_n} \right),
\end{IEEEeqnarray}
where the last inequality is derived from the relation 
\begin{equation}
P_{\overline{X}^n} ({\bf x}_{i_0})\Pr\{X^n \in S_n\} =  P_{{X}^n} ({\bf x}_{i_0}) \ge \frac{1}{M_n}.
\end{equation}

Therefore, noting that $ P_{\overline{X}^n} ({\bf x}_{i_0}) \le  P_{\tilde{X}^n} ({\bf x}_{i_0})$, from (\ref{eq:a-4}) and (\ref{eq:a-5}) we have
\begin{IEEEeqnarray}{rCl}
D_f\left( {X^n}||\tilde{X}^n \right) & \!\le\! & \sum_{i=1}^{i_0-1} P_{\tilde{X}^n} ({\bf x}_i) f\left( \Pr\{X^n \in S_n\} \right) \nonumber \\
&& \!+ \! P_{\tilde{X}^n} ({\bf x}_{i_0}) f\left( \Pr\{X^n \in S_n\} \!-\! e^{\!-n\gamma} \right)\! + e^{-n\gamma }f\left( \frac{1}{M_n} \right) \nonumber \\
& \!\le\! & f\left( \Pr\{X^n \in S_n\} - e^{-n\gamma} \right) + e^{-n\gamma }f\left( \frac{1}{M_n} \right).
\end{IEEEeqnarray}
This completes the proof.
\end{IEEEproof}
\begin{remark}
In the direct part of the proof of Theorem \ref{theo:1-3}, Han has used the pair of mapping which is essentially same with the optimum fixed-length source code.
One may wonder whether we can use a pair of mappings such as 
\begin{equation}
\varphi'_n({\bf x}_i) =\left\{ \begin{array}{cc}  
i  &  {\bf x} \in S_n  \\
1 & \mbox{otherwise} 
\end{array}\right. 
\end{equation}
and $\psi'_n(i) = {\bf x}_i$ in the above lemma. Unfortunately, it is not easy to derive the similar bound by using $(\varphi'_n,\psi_n')$ in the case of $f$-divergences.
Therefore, it is essential to consider the pair of mappings $(\varphi_n,\psi_n)$ in the proof of the theorem.
\end{remark}
%
%
%
\begin{lemma} \label{lemma:converse}
Assuming that the function $f$ satisfies conditions C1) and C3), for any pair of mappings $(\varphi_n,\psi_n)$ it holds that
\begin{IEEEeqnarray}{rCl}
D_f\left(X^n|| \psi_n(\varphi_n(X^n))\right)
&  \!\ge \! & f\left( \Pr \left\{ \frac{1}{n} \log \frac{1}{P_{X^n}(X^n)} \! \le \! \frac{1}{n}\log M_n \! + \! \gamma  \right\} \! + \! e^{\! -n\gamma} \right), \ \ 
\end{IEEEeqnarray}
for any $\gamma >0$.
\end{lemma}
\begin{IEEEproof}
We fix an arbitrary pair of mappings $(\varphi_n,\psi_n)$ and define the probability distribution $P_{\tilde{X}^n}$ by
$
\tilde{X}^n = \psi_n(\varphi_n(X^n)).
$
Define a set $S'_n$ as
\begin{equation}
S'_n := \left\{ {\bf x} \in {\cal X}^n \left| \frac{1}{n} \log \frac{1}{P_{X^n}({\bf x})} \le \frac{1}{n}\log M_n + \gamma \right. \right\}. 
\end{equation}
Then, for $\forall {\bf x} \in (S'_n)^c$ it holds that
\begin{equation}
\frac{1}{n} \log \frac{1}{P_{X^n}({\bf x})} > \frac{1}{n}\log M_n + \gamma.
\end{equation}
Next, we define another set $B_n$ as
\begin{equation}
B_n := \left\{ {\bf x} \in {\cal X}^n |  P_{\tilde{X}^n}({\bf x}) > 0\right\}, 
\end{equation}
and index the element of $B_n$ as 
$
B_n = \{ {\bf x}_1, {\bf x}_2, \dots, {\bf x}_{|B_n|} \}.
$
Then, from the property of mappings $(\varphi_n,\psi_n)$, we obtain
\begin{align} \label{eq:Ln}
|B_n| \le M_n.
\end{align}

Thus, from the condition C3) we obtain 
\begin{IEEEeqnarray}{rCl}
D_f\left({X^n}|| \tilde{X}^n \right)
& = & \sum_{{\bf x} \in B_n} P_{\tilde{X}^n} ({\bf x}) f\left( \frac{P_{{X}^n} ({\bf x})}{P_{\tilde{X}^n} ({\bf x})} \right) \nonumber \\
& \ge & f\left(\Pr \left\{ X^n \in B_n \cap S'_n \right\} + \Pr \left\{ X^n \in B_n \cap (S'_n)^c \right\} \right) \nonumber \\
& \ge & f\left(\Pr \left\{ X^n \in S'_n \right\} + \sum_{{\bf x} \in B_n \cap ({S'_n})^c} P_{X^n}({\bf x})\right) \nonumber \\
& \ge & f\left(\Pr \left\{ X^n \in S'_n \right\} + \sum_{{\bf x} \in B_n \cap ({S'_n})^c} \frac{e^{-n\gamma}}{M_n}\right) \nonumber \\
& \ge & f\left(\Pr \left\{ X^n \in S'_n \right\} + e^{ - n\gamma}\right),
\end{IEEEeqnarray}
for sufficiently large $n$, where the first inequality is due to (\ref{eq:log-sum}) and the second inequality is due to C1), the third inequality is from the definition of $S_n'$, and the last inequality is due to (\ref{eq:Ln}) and  C1).
This completes the proof of the lemma.
\end{IEEEproof}
These lemmas demonstrate the relationship between $f$-divergence and rates of mapping. As will be evident in the following chapter, they are effective for deriving the optimum SRNG rate.
It should be emphasized that in the proof of Lemmas \ref{lemma:direct} and \ref{lemma:converse}, we do not use C2).

\section{Optimum SRNG Rate}
\subsection{General formula}
In this section, first we show the general formula of the SRNG problem with respect to the $f$-divergence by using two lemmas. Second, we apply our general formula to the specified function $f$.

The optimum SRNG rate with respect to the given $f$-divergence is defined as follows.
\begin{definition}
Rate $R$ is said to be $\Delta$-achievable with the given $f$-divergence if there exists a sequence of mappings $(\varphi_n,\psi_n)$ such that 
\begin{equation}
\limsup_{n \rightarrow \infty} D_f\left(X^n||\tilde{X}^n\right) \le \Delta \mbox{ and } \limsup_{n \rightarrow \infty}  \frac{1}{n} \log M_n \leq R,
\end{equation}
where $\tilde{X}^n=\psi_n(\varphi_n(X^n))$.
\end{definition}
\begin{definition}[Optimum SRNG Rate]
\begin{equation}
S_{f}(\Delta|{\bf X}) =  \inf \left\{ R | \mbox{$R$ is $\Delta$-achievable with the given $f$-divergence} \right\}.
\end{equation}
\end{definition}

To express the general formula of the optimum SRNG rate, we define the following quantity which depends on the function $f$.
\begin{IEEEeqnarray}{rCl} \label{eq:K}
K_f(\Delta|{\bf X})
:= \inf \left\{  R \left| \limsup_{n \to \infty } f\left( \Pr \left\{ \frac{1}{n} \log \frac{1}{P_{X^n}(X^n)}  \le  R  \right\} \right) \le \Delta \right. \right\}.
\end{IEEEeqnarray}
Then, we have the following theorem.
\begin{theorem} \label{theo:4-1}
Assuming that the function $f$ satisfies conditions C1)--C3), then for any $0 \le \Delta  < f(0)$ it holds that
\begin{IEEEeqnarray}{rCl}
 S_f(\Delta|{\bf X}) =  {K}_{f}(\Delta|{\bf X}).
\end{IEEEeqnarray}
\end{theorem}
\begin{IEEEproof}
The proof consists of two parts.

(Direct Part:) 
Setting $R={K}_{f}(\Delta|{\bf X})$, we shall show that $R+2\gamma$ is $\Delta$-achievable with the given $f$-divergence for any $\gamma >0$.
To do so, we set $M_n=e^{n(R+2\gamma)}$. Cleary, we have 
\begin{equation}  \label{eq:4-40}
\limsup_{n \to \infty}\frac{1}{n} \log M_n \le R.
\end{equation}
Next, we evaluate the $f$-divergence. From Lemma \ref{lemma:direct}, for the given function $f$ there exists a pair of mappings $(\varphi_n,\psi_n)$ such that
\begin{IEEEeqnarray}{rCl}  \label{eq:a-60}
D_f(X^n||\tilde{X}^n)
&  \le  & 
f\left( \Pr \left\{ \frac{1}{n} \log \frac{1}{P_{X^n}(X^n)}  \le  \frac{1}{n}\log M_n  - \gamma  \right\}  -   e^{- n\gamma} \right) + e^{-n\gamma}f\left(\frac{1}{M_n} \right).
\end{IEEEeqnarray}
Thus, noting that the second term on the RHS in (\ref{eq:a-60}) goes to $0$ because of C2), from the continuity of the function $f$ we have
\begin{IEEEeqnarray}{rCl}
\limsup_{n \to \infty}D_f(X^n||\tilde{X}^n)
 & \le & \limsup_{n \to \infty}f\left( \Pr \left\{ \frac{1}{n} \log \frac{1}{P_{X^n}(X^n)} \le R + \gamma  \right\}  \right) \nonumber \\
& \le & \Delta,
\end{IEEEeqnarray}
where the last inequality is due to the definition of $K_f(\Delta|{\bf X})$, which completes the proof of the direct part.

(Converse Part:)
Suppose that $R$ is $\Delta$-achievable with the given $f$-divergence, then there exists a pair of mapping $(\varphi_n,\psi_n)$ such that
\begin{IEEEeqnarray}{rCl}  \label{eq:r-cond1}
\limsup_{n \to \infty} D_f\left(X^n||\tilde{X}^n\right) & \le & \Delta, \\ \label{eq:r-cond3}
\limsup_{n \to \infty} \frac{1}{n}\log M_n & \le & R,
\end{IEEEeqnarray}
where $
\tilde{X}^n = \psi_n(\varphi_n(X^n))
$.
We fix this pair $(\varphi_n,\psi_n)$.

Then, from Lemma \ref{lemma:converse} and (\ref{eq:r-cond1}) we have
\begin{IEEEeqnarray}{rCl}
\Delta  & \ge & \limsup_{n \to \infty} D_f\left({X^n}||\tilde{X}^n \right) \nonumber \\
& = & \limsup_{n \to \infty} f\left(\Pr \left\{ \frac{1}{n} \log \frac{1}{P_{X^n}(X^n)} \le \frac{1}{n}\log M_n + \gamma  \right\}\right), 
\end{IEEEeqnarray}
where we use the continuity of the function $f$.
Here, from (\ref{eq:r-cond3}) 
\begin{equation}
\frac{1}{n} \log M_n \le R + \gamma,
\end{equation}
for sufficiently large $n$.

Therefore, for the $\Delta$-achievable rate $R$ it must holds that
\begin{IEEEeqnarray}{rCl}
\limsup_{n \to \infty} f\left(\Pr \left\{  \frac{1}{n} \log \frac{1}{P_{X^n}(X^n)} \le R + 2\gamma \right\} \right) \le \Delta.
\end{IEEEeqnarray}
This inequality means that the converse part holds. 
\end{IEEEproof}
\begin{remark}
It should be emphasized that the optimum \textit{resolvability} rate with respect to $f$-divergences has also been characterized by (\ref{eq:K}) \cite[Theorem 3.1]{Nomura_TIT2020}. Hence, the above theorem has shown a kind of similarity between the resolvability and  SRNG problems with a class of $f$-divergences. 
\end{remark}

Next, we show another expression of $S_f(\Delta|{\bf X})$.
We define 
\begin{equation}
f^{-1}(T) = \min \{ t | f(t)=T  \}.
\end{equation}
Furthermore, we introduce the quantity given $0 \le \varepsilon < 1$ so as to express the optimum SRNG rate.
\begin{definition}[$\varepsilon$-spectral sup-entropy rate]
\begin{IEEEeqnarray}{rCl} \label{def:entropy}
\overline{H}(\varepsilon|{\bf X}) \!:=\! &\inf \left\{  R \left| \limsup_{n \to \infty } \Pr \left\{ \frac{1}{n} \log \frac{1}{P_{X^n}(X^n)}  \!>\!  R  \right\} \!\le\! \varepsilon \right. \right\}.
\end{IEEEeqnarray}
\end{definition}
It is known that the optimum $\varepsilon$-fixed-length source coding rate is characterized by $\overline{H}(\varepsilon|{\bf X})$ \cite[Theorem 1.6.1]{Han}.

Then, we have the following theorem.
\begin{theorem}
Assuming that the function $f$ satisfies C1)--C3), it holds that
\begin{IEEEeqnarray}{rCl}
S_f(\Delta|{\bf X}) & = & \overline{H}(1 - f^{-1}(\Delta)|{\bf X}), 
\end{IEEEeqnarray}
\end{theorem}
\begin{IEEEproof}
It is clear from Theorems \ref{theo:4-1}, and the definition of $f^{-1}$.
\end{IEEEproof}

\subsection{Particulalization}
In this subsection, we focus on the specified function $f$ which satisfies conditions C1)--C3) and 
compute $S_f(\Delta|{\bf X})$ by using Theorem \ref{theo:4-1}. 
We use the notation
$D_f(X^n||\tilde{X}^n) := D_f(X^n||\psi_n(\varphi_n(X^n)))$
for short. 

\subsubsection{Variational distance}
We first consider the case of $f(t) = (1-t)^+$, which indicates the variational distance.
From Theorem \ref{theo:4-1} we obtain the following corollary:
\begin{corollary} 
\label{coro:vd}
For $f(t) = (1-t)^+$, it holds that
\begin{IEEEeqnarray}{rCl}
{S}_{f}(\Delta|{\bf X}) &=&  \overline{H}(\Delta|{\bf X}).
\end{IEEEeqnarray} 
\end{corollary}
\begin{IEEEproof}
For $f(t) = (1-t)^+$, we clearly have
\begin{IEEEeqnarray}{rCl}
{K}_f(\Delta|{\bf X}) & = &  \inf \left\{  R \left| \limsup_{n \to \infty } \left( 1 \!-\!\Pr \left\{ \frac{1}{n} \log \frac{1}{P_{X^n}(X^n)}  \!\le\!  R  \right\}\right)^+ \!\le\! \Delta  \right. \right\} \nonumber \\
& = &  \inf \left\{  R \left| \limsup_{n \to \infty } \Pr \left\{ \frac{1}{n} \log \frac{1}{P_{X^n}(X^n)}  >  R  \right\} \le \Delta \right. \right\}.
\end{IEEEeqnarray}
Hence, we obtain the corollary.
\end{IEEEproof}
When $\Delta=0$, the above corollary coincides with Theorem \ref{theo:1-3}. Thus, the above result is a generalization of Theorem \ref{theo:1-3}.

Similarly to the derivation of the above corollary, we immediately obtain the optimum SRNG rate with other approximation measures by calculating the $f^{-1}(t)$. Hence, we omit these proofs.
%
%
\subsubsection{Reverse KL divergence}
We consider the case of $f(t) = -\log t$, which indicates
$D_f(X^n||\tilde{X}^n) 
=  \sum_{{\bf x} \in {\cal X}^n} P_{\tilde{X}^n}({\bf x}) \log \frac{P_{\tilde{X}^n}({\bf x})}{P_{{X}^n}({\bf x})}.$
We obtain the corollary:
\begin{corollary} 
For $f(t) = -\log t$, it holds that
\begin{IEEEeqnarray}{rCl}
{S}_{f}(\Delta|{\bf X}) &=&  \overline{H}(1 - e^{-\Delta}|{\bf X}).
\end{IEEEeqnarray}
\end{corollary}
%
%
\subsubsection{Hellinger distance}
We consider the case of $f(t) = 1 - \sqrt{t}$, which indicates the Hellinger distance.
\begin{corollary} 
For $f(t) =  1-\sqrt{t}$, it holds that
\begin{IEEEeqnarray}{rCl}
{S}_{f}(\Delta|{\bf X}) &=&  \overline{H}(2\Delta-\Delta^2|{\bf X}).
\end{IEEEeqnarray} 
\end{corollary}
\subsubsection{$E_\gamma$-divergence}
Finally, we consider the case that $f(t) = (\gamma-t)^+ + 1 - \gamma$, which indicates the $E_\gamma$-divergence.
\begin{corollary} 
\label{coro:egamma}
For $f(t) =  (\gamma-t)^+ + 1 - \gamma$, it holds that
\begin{IEEEeqnarray}{rCl}
{S}_f({\Delta}|{\bf X}) &=&  \overline{H}(\Delta|{\bf X}).
\end{IEEEeqnarray}
\end{corollary}
The above corollary shows that the optimum SRNG rate with respect to the $E_\gamma$-divergence does not depend on $\gamma$, which implies that it coincides with the optimum SRNG rate with respect to the variational distance (cf. Corollary \ref{coro:vd}).
%
%
%
\subsection{Discussion}
We next consider a relationship to other typical problems in information theory, such as the fixed-length source coding and the source resolvability.

As we have mentioned, the optimum $\varepsilon$-fixed-length source coding rate is characterized by $\overline{H}(\varepsilon|{\bf X})$.
We first define the achievability in the fixed-length source coding problem.
Let $\phi_n: {\cal X}^n \to {{\cal M}_n}$, $\xi_n: {\cal M}_n \to {\cal X}^n$ be a pair of  fixed-length source coding encoder and decoder, respectively, for the source ${\bf X} = \{ X^n \}_{n=1}^\infty$. 
The decoding error probability $\varepsilon_n$ is defined by 
$
\varepsilon_n \equiv \Pr \left\{ X^n \neq \psi_n(\varphi_n(X^n)) \right\}.
$
Such a code is denoted by $(n, M_n, \varepsilon_n)$.
\begin{definition}
Rate $R$ is said to be $\varepsilon$-achievable if there exists a code $(n, M_n, \varepsilon_n)$ such that 
\begin{equation}
\limsup_{n \rightarrow \infty} \varepsilon_n \leq \varepsilon \mbox{ and } \limsup_{n \rightarrow \infty}  \frac{1}{n} \log M_n \leq R.
\end{equation}
\end{definition}
\begin{definition}[$\varepsilon$-fixed-length source coding rate]
\begin{equation}
R(\varepsilon|{\bf X}) =  \inf \left\{ R | \mbox{$R$ is $\varepsilon$-achievable} \right\}.
\end{equation}
\end{definition}

Then, we have
\begin{theorem}[Steinberg and Verd\'{u} \cite{Steinberg}, Han \cite{Han}] 
\begin{equation}
R(\varepsilon|{\bf X}) = \overline{H}(\varepsilon|{\bf X}).
\end{equation}
\end{theorem}

The source resolvability problem is also a fundamental challenge in information theoretic random number generation. It involves approximating an arbitrary general source using a discrete uniform random number, with the goal of minimizing the size of this random number.
The optimum resolvability rate with respect to the given $f$-divergence has been investigated in \cite{Nomura_TIT2020}. Let $U_{M_n}$ denote the random variable distributed on $\{1,2,\dots,M_n\}$ uniformly.

\begin{definition}
Rate $R$ is said to be $\delta$-achievable if there exists a mapping $\phi_n : {{\cal M}_n} \to {\cal X}^n$ with the given $f$-divergence such that 
\begin{equation}
\limsup_{n \rightarrow \infty} \frac{1}{n} \log M_n \leq R \mbox{ and } \limsup_{n \rightarrow \infty} D_f(X^n||\phi_n(U_{M_n})) \leq \delta.
\end{equation}
\end{definition}
\begin{definition} 
\begin{eqnarray} 
L_{f}(\delta|{\bf X}) = \inf \left\{ R \left|R \mbox{ is $\delta$-achievable with the given $f$-divergence} \right. \right\}.
\end{eqnarray}
\end{definition}
Then, from the result given in \cite{Nomura_TIT2020}, Theorem \ref{theo:1-3} and Theorem \ref{theo:4-1}, we obtan the following corollary which reveals a deep relationship between the fixed-length source coding problem, the resolvability problem and the SRNG problem.
\begin{corollary}  \label{coro:4-3}
Assuming that the function $f$ satisfies conditions C1)--C3), then for any $0 \le \Delta  < f(0)$ it holds that
\begin{IEEEeqnarray}{rCl}
S_f(\Delta|{\bf X}) = L_f(\Delta|{\bf X})=  R(1 - f^{-1}(\Delta)|{\bf X}).
\end{IEEEeqnarray}
\end{corollary}
%
%
%
%
%
%
%
%
\section{Application to Rate distortion perception problem}
Rate-distortion theory highlights the trade-off between information rate and distortion \cite{Shannon,Cover}. Yet, in practical scenarios like image processing, reduced distortion does not invariably lead to enhanced perceptual quality.
Blau and Michaeli conceptualized the perceptual quality of reconstructed information in terms of the variational distance between the probability distribution of the source and that of the reconstructed information \cite{BM2019}, demonstrating a \lq\lq perception-distortion tradeoff".
Matsumoto has first attempted to incorporate this perceptional quality into the rate-distortion theory \cite{Matsumoto2018,Matsumoto2019}. He has introduced the tradeoff among three quantities: the information rate, distortion and perceptual quality. He has also derived the general formula of the rate distortion dispersion (RDP) function.
In this work, we extend this framework by employing $f$-divergence as the criterion for perceptual quality, rather than the variational distance.
For simplicity, we assume that the reconstruction alphabet ${\cal Y}^n $ is a subset of ${\cal X}^n$.

Let $\phi_n: {\cal X}^n \to {\cal M}_n$ and $\xi_n: {\cal M}_n \to {\cal X}^n$ denote a fixed-length lossy source encoder and a decoder, respectively. A general distortion function is defined by a mapping $g_n: {\cal X}^n \times {\cal X}^n \to [0, + \infty)$, where $g_n({\bf x}, {\bf x})=0$ for any ${\bf x} \in {\cal X}^n$.

We define the RDP function with respect to $f$-divergences.
\begin{definition}  \label{def:rdp}
A triplet $(R,D,\Delta)$ is said to be achievable with  the given $f$-divergence if there exists a sequence of $(\phi_n,\xi_n)$ such that
\begin{IEEEeqnarray}{rCl}
\limsup_{n \rightarrow \infty}  \frac{1}{n} \log M_n & \leq & R,  \\ \label{eq:70}
\limsup_{n \rightarrow \infty} \frac{1}{n} \mathbb{E}\left[g_n\left(X^n,\xi_n(\phi_n(X^n))\right)\right] & \le & D,  \\ \label{eq:fcond}
\limsup_{n \to \infty}D_f(X^n||\xi_n(\phi_n(X^n))) & \le & \Delta.
\end{IEEEeqnarray}
\end{definition}
\begin{definition}[RDP function with the given $f$-divergence]
\begin{equation}
R_f(D,\Delta) =  \inf \left\{ R | \mbox{$(R,D,\Delta)$ is achievable with the given $f$-divergence} \right\}.
\end{equation}
\end{definition}
Then, from Theorem \ref{theo:4-1} we immediately have the following theorem.
\begin{theorem}  \label{theo:5-1}
\begin{equation}
R_f(D,\Delta) \ge \max \{ r(D|{\bf X}), K_f(\Delta|{\bf X}) \},
\end{equation}
where $r(D|{\bf X})$ is the general RD function (see, \cite{Steinberg,Han}) and $K_f(\Delta|{\bf X})$ is defined in (\ref{eq:K}).
\end{theorem}

\begin{IEEEproof}
The theorem is obvious from Theorem \ref{theo:4-1} and the result about the RD function in general setting \cite{Steinberg,Han}.
\end{IEEEproof}
The above theorem shows the lower bound of the RDP function with the given $f$-divergence.
Deriving the upper bound in the general case, however, presents significant challenges.
To facilitate this in a special case, we define two key quantities: $\overline{g_n}:= \max_{({\bf x},{\bf x}')}g_n({\bf x},{\bf x}') $ and $D_{threshold}:= \frac{1}{n}\overline{g_n} \cdot \Pr\{- \log P_{X^n}(X^n) \ge K_f(\Delta|{\bf X}) \}$ , where $\Delta$, $f$, and $g_n$ are given.
\begin{theorem}  \label{theo:5-2}
For $D \ge D_{threshold}$
\begin{equation}
R_f(D,\Delta) \le K_f(\Delta|{\bf X}) .
\end{equation}
\end{theorem}
\begin{IEEEproof}
We can prove this theorem by using the pair of mappings $(\varphi_n,\psi_n)$ used in the proof of Lemma \ref{lemma:direct}.
If we employ $(\varphi_n,\psi_n)$ as the lossy source encoder and the decoder, then for all ${\bf x} \in S_n \cup T_n$ it holds that ${\bf x} = \psi_n(\varphi_n({\bf x}))$. This yields 
$g_n({\bf x},\psi_n(\varphi_n({\bf x})))=0$ for ${\bf x} \in S_n \cup T_n$.
Hence, if use this pair of mappings with 
\begin{equation}
\frac{1}{n} \log M_n = K_f(\Delta|{\bf X}) + \gamma
\end{equation}
as the lossy source coding encoder and the decoder, then we obtain
\begin{IEEEeqnarray}{rCl}
\frac{1}{n}\mathbb{E}\left[ g_n(X^n,\psi_n(\varphi_n(X^n)))\right] & = & \frac{1}{n} \sum_{{\bf x} \notin S_n \cup T_n} P_{X^n}({\bf x})g_n(X^n,\psi_n(\varphi_n(X^n))) \nonumber \\
& \le & \frac{1}{n}\overline{g_n} \cdot \Pr\{- \log P_{X^n}(X^n) \ge K_f(\Delta|{\bf X}) \} \nonumber \\
& = & D_{threshold}.
\end{IEEEeqnarray}
Therefore, if $D \ge D_{threshold}$ holds, the constraints regarding distortion can be considered negligible.
This means that the theorem holds.
\end{IEEEproof}
Intuitively, relaxing the constraints on distortion levels leads to the prominence of the condition stated in (12) regarding perceptual quantity.
This phenomenon is highlighted by the condition $D \ge D_{threshold}$ in the above theorem.

Originally, the RDP function has been defined with respect to the variational distance instead of the $f$-divergence in (\ref{eq:fcond}) \cite{Matsumoto2019}.
Matsumoto has shown the following theorem.
\begin{theorem}[RDP function\cite{Matsumoto2019}]  \label{theo:M}
For $f(t) = (1-t)^+$, it holds that 
\begin{equation}
R_f(D,\Delta) = \max \{ r(D|{\bf X}), \overline{H}(\Delta|{\bf X}) \},
\end{equation}
where $\overline{H}(\Delta|{\bf X})$ is defined in (\ref{def:entropy}).
\end{theorem}
This theorem demonstrates that for a special case of $f$-divergence, specifically when using the variational distance, the upper and lower bounds coincide. This is due to the close relationship between the variational distance and the error probability in coding.
On the other hand, it remains challenging to show similar results when employing general $f$-divergences.
\begin{remark}
In Def. \ref{def:rdp}, a constraint on average distortion (\ref{eq:70}) has been imposed. On the other hand, it is also possible to define the RDP function on the basis of the maximum distortion. In this case, using Theorem \ref{theo:4-1}, a theorem analogous to Theorem \ref{theo:5-1} can be derived.
\end{remark}

It is noteworthy that Theis and Wagner have explored the RDP function in a more general context \cite{LW2021}, presenting an approach that diverges from the one discussed in this paper. Deriving the upper bound (Direct Part) of the RDP function with $
f$-divergence generally poses significant challenges. However, we posit that Lemma \ref{lemma:direct} and its accompanying proof offer valuable insights for constructing effective source codes in the rate-distortion perception problem, highlighting an advantage of our approach.%
\section{Alternative Expression of Optimum SRNG Rate using Smooth R\'enyi entropy}
It is known that the optimum $\varepsilon$-fixed-length source coding rate is characterized by using the smooth R\'enyi entropy of the source \cite{Uyematsu_ISIT2010,Uyematsu2010}.
The result in \cite{Uyematsu2010} together with Corollary \ref{coro:4-3} implies that the optimum SRNG rate has also been characterized by the smooth R\'enyi entropy of the source.

In this section, we try to express the optimum SRNG rate with $f$-divergences by using the smooth max entropy of the source.
\begin{definition}[Smooth R\'enyi entropy of order $\alpha$ \cite{RW04}]
The smooth R\'enyi entropy of order $\alpha$ given $\delta \ (0 \le \delta <1)$ is defined by
\begin{equation} \label{eq:renyi}
H_{\alpha}(\delta| {X^n}) := \frac{1}{1 - \alpha} \inf_{ P_{\overline{X}^n} \in B^\delta(P_{X^n})} \log \left( \sum_{{\bf x} \in {\cal X}^n} P_{\overline{X}^n}({\bf x})^\alpha  \right),
\end{equation}
where
\begin{IEEEeqnarray}{rCl} 
B^\delta(P_{X^n})
& := & \left\{ P_{\overline{X}^n} \in {\cal P}^n \left| \frac{1}{2} \sum_{{\bf x} \in {\cal X}^n} |P_{X^n}({\bf x}) - P_{\overline{X}^n}({\bf x})  |  \le \delta \right. \right\}.\ \ 
\end{IEEEeqnarray}
\end{definition}
The smooth R\'enyi entropy of order $0$ is called the smooth max entropy. The following theorem given by Uyematsu \cite{Uyematsu_ISIT2010,Uyematsu2010} has shown another expression of the smooth max entropy.
\begin{theorem}[Uyematsu \cite{Uyematsu_ISIT2010,Uyematsu2010} ]
\begin{equation}
H_{0}(\delta| {X^n}) = \min_{\substack{A_n \subset {\cal X}^n \\ \Pr\{X^n \in A_n \} \ge 1 - \delta  }} \log |A_n|.
\end{equation}
\IEEEQED
\end{theorem}
In this section, we use the above expression of the smooth max entropy instead of (\ref{eq:renyi}).
We first introduce two fundamental lemmas. Proofs of these lemmas are given in Appendices. In this section, we impose the following assumption.
\begin{equation} \label{eq:assump_source}
\overline{H}({\bf X}) < +\infty,
\end{equation}
where 
\begin{equation}
\overline{H}({\bf X}) := \inf\left\{ R \left| \lim_{n \to \infty} \Pr \left\{ \frac{1}{n} \log \frac{1}{P_{X^n}(X^n)} \le R \right\} =1 \right. \right\}.
\end{equation}

Then, we have
\begin{lemma} \label{lemma:direct2}
Assuming that the function $f$ satisfies C1)--C3), for any  $\gamma >0$ and any $M_n$ satisfying
\begin{equation} \label{eq:assump1}
\frac{1}{n} \log M_n \ge \frac{1}{n}H_0(1 - f^{-1}(\Delta)|X^n) + \gamma,
\end{equation}
there exists a pair of mapping $(\varphi_n,\psi_n)$ which satisfies
\begin{equation} \label{eq:theorem_1-2}
D_f(X^n||\psi_n(\varphi_n(X^n))) \le \Delta + \gamma,
\end{equation}
for sufficiently large $n$.
\end{lemma}

\begin{lemma} \label{lemma:converse2}
Assuming that the function $f$ satisfies C1) and C3), for any pair of mappings $(\varphi_n,\psi_n)$ satisfying
\begin{equation}
D_f(X^n||\psi_n(\varphi_n(X^n))) \le \Delta,
\end{equation}
it holds that
\begin{equation}
\frac{1}{n} \log M_n \ge \frac{1}{n}H_0(1 - f^{-1}(\Delta)|X^n).
\end{equation}
\end{lemma}

The following theorem shows an another expression of the optimum SRNG problem.
\begin{theorem} \label{theo:5-3}
Under conditions C1)--C3), for any $0 \le \Delta <f(0)$, it holds that
\begin{align}
S_f(\Delta|{\bf X}) & = \lim_{\nu \downarrow 0}\limsup_{n \to \infty} \frac{1}{n}H_{0}(1 - f^{-1}(\Delta+\nu)|X^n) \nonumber \\
& = \lim_{\nu \downarrow 0}\limsup_{n \to \infty} \frac{1}{n} H_{0}(1 - f^{-1}(\Delta) + \nu|X^n).
\end{align}
\end{theorem}
The optimum $\varepsilon$-fixed-length-source coding rate is also characterized by using $H_0(\varepsilon|X^n)$ \cite{Uyematsu2010}. Thus, from the result in \cite{Uyematsu2010} together with Corollary \ref{coro:4-3}, we are able to obtain the above theorem. However, we describe whole proofs of Lemmas \ref{lemma:direct2} and \ref{lemma:converse2}, and Theorem \ref{theo:5-3} in appndices so as to reveal the fundamental logic underlying the process of the SRNG problem.

In the proof of Lemma \ref{lemma:direct2}, we consider a set ${\cal C}_n$ of sequences with high probability, and contemplate a mapping where sequences belonging to set $({\cal C}_n)^c$ correspond to those in set ${\cal C}_n$. In particular, the mapping is adjusted to ensure that the probability of sequences belonging to ${\cal C}_n$ does not become excessively high after two mappings. This construction is essentially similar to mappings used in the proof of Lemma \ref{lemma:direct}. Therefore, in the SRNG problem, it is considered essential to map the sequences of the lower probability set $({\cal C}_n)^c$ to set ${\cal C}_n$ in a well-balanced manner.
Lemmas \ref{lemma:direct} and \ref{lemma:direct2} demonstrate the construction methods for this well-balanced mapping.
\section{Concluding Remarks}
This paper addresses the SRNG problem in the context of $f$-divergences.
To derive the optimum SRNG rate with respect to the given $f$-divergence, we initially established two finite-blocklength bounds.
Subsequently, we presented the general formula for the optimum SRNG rate, incorporating the function $f$ and the information spectrum quantity.
The results reveal a form of duality between the problems of source coding and self-random number generation.

We have extended our general formula to the RDP problem, deriving a lower bound of the RDP function. This constitutes a generalization of Matsumoto's findings \cite{Matsumoto2019}.
Matsumoto has demonstrated the RDP function in relation to the variational distance by integrating the optimal resolvability rate \cite{HV93} with the RD function \cite{Matsumoto2019}.
Similarly, Theorem \ref{theo:5-1} can be derived by amalgamating the optimum resolvability rate with $f$-divergence \cite{Nomura_TIT2020} and the RD function.
While the general formula of the upper bound of the RD function remains unproven in this work, the mapping constructions delineated in the proof of Lemma \ref{lemma:direct} offer valuable insights for addressing the lossy source code within the RDP problem.

The definition of second-order optimum achievable rates in the SRNG problem parallels the approach outlined in previous studies \cite{Hayashi, Hayashi2, NH2011, YHN}.
For second-order analysis, we effectively utilize Lemmas \ref{lemma:direct} and \ref{lemma:converse}.
%
%


\ifCLASSOPTIONcaptionsoff
  \newpage
\fi



%

\newpage

\appendices
\section{Proof of Lemma \ref{lemma:direct2}}  \label{app1}
\renewcommand{\theequation}{A.\arabic{equation}}
We fix $M_n$ satisfying (\ref{eq:assump1}) and show that there exists a pair of mappings $(\varphi_n,\psi_n)$ that satisfies (\ref{eq:theorem_1-2}) for sufficiently large $n$.
We consider a set $B_n \subset {\cal X}^n$ satisfying
\begin{equation} \label{eq:a-0-0}
\Pr\{X^n \in B_n \} \ge  f^{-1}(\Delta)
\end{equation}
and
\begin{equation} \label{eq:b-b-1}
\log |B_n| = H_0(1 - f^{-1}(\Delta)|X^n).
\end{equation}
There may be several sets that satisfy the aforementioned conditions. In that case, we choose $B_n$ such that for any ${\bf x} \in B_n$ and ${\bf x}' \notin B_n$,  $P_{ X^n}({\bf x}) \ge P_{ X^n}({\bf x}')$ holds.

We arrange elements in ${\cal X}^n$ as
$
{\cal X}^n = \{{\bf x}_1,{\bf x}_2,\dots, \}
$
according to $P_{{X}^n}({\bf x})$ in descendant order.
Since the inequality
\begin{equation}
\frac{|B_n|}{M_n} \le e^{-n\gamma}
\end{equation}
holds, 
$\{{\bf x}_1,{\bf x}_2,\dots, {\bf x}_{|B_n|},\dots,{\bf x}_{M_n},\dots\}$
holds for sufficiently large $n$.

In addition, we set ${\cal C}_n$ as 
\begin{equation}
{\cal C}_n := \{{\bf x}_1,{\bf x}_2,\dots, {\bf x}_{|B_n|},\dots,{\bf x}_{M_n}\}.
\end{equation}
Then, 
\begin{equation} \label{eq:1/M}
P_{X^n}({\bf x}) \le \frac{1}{M_n}
\end{equation}
holds for each ${\bf x} \notin {\cal C}_n$.

We define the probability distribution $P_{\overline{X}^n}$ over $B_n$ as
\begin{equation}
P_{\overline{X}^n}({\bf x}) := \left\{ \begin{array}{cc}
\frac{P_{{X}^n}({\bf x})}{\Pr\{ X^n \in B_n \}} & {\bf x} \in B_n, \\
0 & \mbox{otherwise}.
\end{array} \right.
\end{equation}

To construct a mapping, we assign a set of sequence $A_n(i) \subseteq ({\cal C}_n)^c$
for each ${\bf x}_i \in B_n$ according to the following procedure.

For ${\bf x}_1 \in B_n$ we assign $A_n(1) \subseteq ({\cal C}_n)^c$ that satisfies
\begin{equation}
P_{{X}^n}({\bf x}_1) + \sum_{{\bf x} \in A_n(1)}P_{{X}^n}({\bf x})  \le P_{\overline{X}^n}({\bf x}_1)
\end{equation}
and 
\begin{equation}
P_{{X}^n}({\bf x}_1) + \sum_{{\bf x} \in A_n(1)}P_{{X}^n}({\bf x}) + P_{{X}^n}({\bf x}')  > P_{\overline{X}^n}({\bf x}_1)
\end{equation}
where ${\bf x}'$ is any sequence in $({\cal C}_n)^c \setminus A_n(1)$.

Similarly, for ${\bf x}_2 \in B_n$ we assign $A_n(2) \subseteq ({\cal C}_n)^c$ that satisfies
\begin{equation}
P_{{X}^n}({\bf x}_2) + \sum_{{\bf x} \in A_n(2)}P_{{X}^n}({\bf x})  \le P_{\overline{X}^n}({\bf x}_2)
\end{equation}
and 
\begin{equation}
P_{{X}^n}({\bf x}_2) + \sum_{{\bf x} \in A_n(2)}P_{{X}^n}({\bf x}) + P_{{X}^n}({\bf x}')  > P_{\overline{X}^n}({\bf x}_2)
\end{equation}
where ${\bf x}'$ is an any sequence in $({\cal C}_n)^c \setminus (A_n(1) \cup A_n(2) )$.

In the similar way, we repeat this operation to choose $A_n(i)$ for ${\bf x}_i$ as long as possible. Suppose that this operation stops at $i_0$. Then, we set 
$A_n(i_0) = ({\cal C}_n)^c \setminus \bigcup_{i=1}^{i_0-1} A_n(i)$.
Since $\Pr\{ X^n \in (B_n \cup {\cal C}^c_n) \} \le 1$ holds, $i_0$ is smaller than or equalt to $|B_n|$.
If $i_0 < |B_n|$ holds, then we set $A_n(j) = \emptyset$ for $i_0 < j \le |B_n|$.
Furthermore, for ${\bf x}_{|B_n|+j} \ (1 \le \forall j \le M_n-|B_n|) $ we also set $A_n({|B_n|+j}) = \emptyset$.

Using these set $A_n(i)$, we define 
a pair of mappings
$\varphi_n: {\cal X}^n \to {\cal M}_n, \quad \psi_n: {\cal M}_n \to {\cal X}^n$
as follows
\begin{equation}
\varphi_n({\bf x}) = i, \mbox{ for }  {\bf x} \in \left( \{{\bf x}_i\} \cup A_n(i) \right),
\end{equation}
\begin{equation}
\psi_n(i)  =  {\bf x}_i.
\end{equation}

We evaluate the performance of the mapping $\phi_n$.
Then, from the construction of the mapping and (\ref{eq:1/M}), for any $i$ satisfying $1 \le  i\le i_0-1$, it holds that
\begin{equation} \label{eq:a-1}
P_{\tilde{X}^n} ({\bf x}_i) \le P_{\overline{X}^n} ({\bf x}_i)  < P_{\tilde{X}^n} ({\bf x}_i) + \frac{1}{M_n},
\end{equation}
On the other hand, we obtain
\begin{align}
P_{\tilde{X}^n} ({\bf x}_{i_0}) - P_{\overline{X}^n} ({\bf x}_{i_0})
& = 1\! - \! \Pr\{X^n \in {\cal C}_n \setminus B_n\} -\sum_{i=1}^{i_0-1}P_{\tilde{X}^n} ({\bf x}_{i}) - \left(1\! - \!\sum_{i=1}^{i_0-1}P_{\overline{X}^n} ({\bf x}_{i})\right) \nonumber \\
& \le \sum_{i=1}^{i_0-1} \left( P_{\overline{X}^n} ({\bf x}_{i}) - P_{\tilde{X}^n} ({\bf x}_{i}) \right) \nonumber \\
& < \frac{|B_n|}{M_n} \le e^{-n\gamma},
\end{align}
where the second inequality is due to (\ref{eq:a-1}).

In addition, for any $i$ satisfying $i_0+1 \le  i \le M_n$, it holds that
\begin{equation} 
P_{\tilde{X}^n} ({\bf x}_{i}) = P_{{X}^n} ({\bf x}_{i}).
\end{equation}

From the above argument the $f$-divergence is given by
\begin{align}   \label{eq:a-42}
D_f\left( {X^n}||\tilde{X}^n\right)
 & = \sum_{i=1}^{M_n} P_{\tilde{X}^n} ({\bf x}_i) f \left(\frac{P_{{X}^n} ({\bf x}_i)}{P_{\tilde{X}^n} ({\bf x}_i)} \right) \nonumber \\
& = \sum_{i=1}^{i_0-1} P_{\tilde{X}^n} ({\bf x}_i) f\left( \frac{P_{{X}^n} ({\bf x}_i)}{P_{\tilde{X}^n} ({\bf x}_i) } \right) + P_{\tilde{X}^n} ({\bf x}_{i_0}) f\left( \frac{P_{{X}^n} ({\bf x}_{i_0})}{P_{\tilde{X}^n} ({\bf x}_{i_0}) } \right) \nonumber \\
& = \sum_{i=1}^{i_0-1} P_{\tilde{X}^n} ({\bf x}_i) f\left( \frac{P_{\overline{X}^n} ({\bf x}_i)\Pr\{X^n \in B_n\}}{P_{\tilde{X}^n} ({\bf x}_i) } \right)  + P_{\tilde{X}^n} ({\bf x}_{i_0}) f\left( \frac{P_{{X}^n} ({\bf x}_{i_0})}{P_{\tilde{X}^n} ({\bf x}_{i_0}) } \right) \nonumber \\
& \le \sum_{i=1}^{i_0-1} P_{\tilde{X}^n} ({\bf x}_i) f\left( \Pr\{X^n \in B_n\} \right) \nonumber \\
& \quad + \left( P_{\overline{X}^n} ({\bf x}_{i_0}) \!+ \! e^{-n\gamma} \right) f\left( \frac{P_{\overline{X}^n} ({\bf x}_{i_0})\Pr\{X^n \in B_n\}}{P_{\overline{X}^n} ({\bf x}_{i_0}) + e^{-n\gamma}} \right),
\end{align}
where the last inequality is due to (\ref{eq:a-1}) and C1).

In order to evaluate the second term of the RHS of (\ref{eq:a-42}), we use the relation
\begin{align} \label{eq:f12}
P_{\overline{X}^n} ({\bf x}_{i_0})
& = ( 1- e^{-n  \gamma  } )P_{\overline{X}^n} ({\bf x}_{i_0}) +e^{-n \gamma }P_{\overline{X}^n}  ({\bf x}_{i_0}).
\end{align}
Then, we have
\begin{align}
\lefteqn{\left( P_{\overline{X}^n} ({\bf x}_{i_0}) + e^{-n\gamma} \right) f\left( \frac{P_{\overline{X}^n} ({\bf x}_{i_0})\Pr\{X^n \in B_n\}}{P_{\overline{X}^n} ({\bf x}_{i_0}) + e^{-n\gamma}} \right)} \nonumber \\
& \le  P_{\overline{X}^n} ({\bf x}_{i_0}) f\left( (1- e^{-n \gamma })  \Pr\{X^n \in B_n\} \right) + e^{-n\gamma} f\left( \frac{e^{-n \gamma } P_{{X}^n} ({\bf x}_{i_0})}{ e^{-n\gamma}} \right) \nonumber \\
& \le P_{\overline{X}^n} ({\bf x}_{i_0}) f\left( (1- e^{-n \gamma })  \Pr\{X^n \in B_n\} \right) + e^{-n\gamma} f\left( e^{-n (\overline{H}({\bf X}) + \gamma)}\right), 
\end{align}

Hence, from C2) and the continuity of the function $f$, for $\forall \nu >0$ we have 
\begin{align} \label{eq:a-4-2}
\lefteqn{\left(P_{\overline{X}^n} ({\bf x}_{i_0}) + e^{-n\gamma}\right) f\left( \frac{P_{\overline{X}^n} ({\bf x}_{i_0}) \Pr\left\{ X^n \in B_n\right\}}{P_{\overline{X}^n} ({\bf x}_{i_0}) + e^{-n\gamma} } \right)} \nonumber \\
& \le  P_{\overline{X}^n} ({\bf x}_{i_0})  f\left(  \Pr\left\{ X^n \in B_n\right\}  - e^{-n (\underline{H}({\bf X}) -\gamma )} \right)   + \nu \nonumber \\
& \le  P_{\overline{X}^n} ({\bf x}_{i_0})  f\left(  \Pr\left\{ X^n \in B_n\right\}\right)   + 2\nu,
\end{align}

Substituting (\ref{eq:a-4-2}) into (\ref{eq:a-42}), we obtain
\begin{align}  \label{eq:a-6}
D_f\left( {X^n}||\psi_n(\varphi_n(X^n)) \right)
 & \le  \sum_{i=1}^{i_0} P_{\tilde{X}^n} ({\bf x}_i) f \left( \Pr \left\{ X^n \in B_n\right\} \right) + 2 \nu \nonumber \\
 & = f \left( \Pr \left\{ X^n \in B_n\right\} \right) +2 \nu \nonumber \\
 & \le f \left( f^{-1}(\Delta) \right) + 2 \nu \nonumber \\
 & = \Delta + 2 \nu,
\end{align}
for sufficiently large $n$ where the last inequality is due to (\ref{eq:a-0-0}).
\IEEEQED

\newpage 

\section{Proof of Lemma \ref{lemma:converse2}}  \label{app2}
\renewcommand{\theequation}{B.\arabic{equation}}

It suffices to show the claim that the relation
\begin{equation} \label{eq:b1}
\frac{1}{n} \log M_n < \frac{1}{n}H_0(1 - f^{-1}(\Delta)|X^n),
\end{equation}
necessarily yields
\begin{equation}
D_f(X^n||\psi_n(\varphi_n(X^n))) > \Delta.
\end{equation}
We denote $H' := H_0(1 - f^{-1}(\Delta)|X^n)$ for short.

For any fixed mapping $ (\varphi_n,\psi_n)$, we set 
$\tilde{X}^n: = \psi_n(\varphi_n(X^n))$ and 
\begin{equation}
B_n := \left\{  {\bf x} \in {\cal X}^n | P_{\tilde{X}^n}({\bf x}) > 0\right\},
\end{equation}
Then, from the property of the mapping it holds that
\begin{equation} \label{eq:MB}
M_n \ge |B_n|.
\end{equation}

From C2), the $f$-divergence is given by
\begin{align}
D_f\left( {X^n}||\tilde{X}^n\right) & = \sum_{{\bf x} \in B_n} P_{\tilde{X}^n}({\bf x}) f\left( \frac{P_{{X}^n}({\bf x})}{P_{\tilde{X}^n}({\bf x})}  \right) \nonumber \\
& \ge  f\left(  \Pr\{ X^n \in B_n  \}  \right)  \nonumber \\
& \ge  f\left(  \max_{ \substack{B_n \subset {\cal{X}}^n \\ |B_n| \le M_n  }}  \Pr\{ X^n \in B_n  \}  \right)  \nonumber \\
& \ge  f\left(  \max_{ \substack{B_n \subset {\cal{X}}^n \\ \log |B_n| < H'  }}  \Pr\{ X^n \in B_n  \}  \right) \nonumber \\
& >  f\left( 1 - ( 1 - f^{-1}(\Delta)) \right) \nonumber \\
& = \Delta,
\end{align}
where the first inequality is due to (\ref{eq:log-sum}), the second inequality is due to (\ref{eq:MB}) and the third inequality is from (\ref{eq:b1}). The last inequality is from the definition of $H'= H_0(1 - f^{-1}(D)|X^n)$. This completes the proof.
\IEEEQED

\newpage

\section{Proof of Theorem \ref{theo:5-3}}  \label{app3}
\renewcommand{\theequation}{C.\arabic{equation}}
We onlly show the first equality, because the second equality can be derived from the first inequality together with the continuity of the function  $f^{-1}$.
The proof consists of two parts.

(Direct Part:)
Fix $\nu>0$ arbitrarily.
From Lemma \ref{lemma:direct2}, for any $\gamma >0$, there exists a pair of mappings $(\varphi_n,\psi_n)$ such that
\begin{IEEEeqnarray}{rCl} \label{eq:t1}
\frac{1}{n} \log M_n
& \le & \frac{1}{n}H_0(1 - f^{-1}(\Delta+\nu)|X^n) +  \gamma, 
\end{IEEEeqnarray}
and
\begin{equation}
D_f(X^n||\psi_n(\varphi_n(X^n))) \le \Delta + \nu + \gamma.
\end{equation}
We here use the diagonal line argument \cite{Han}. Fix a sequence $\{\gamma_i\}_{i=1}^\infty$ such that $\gamma_1 > \gamma_2 > \dots > 0$ and we repeat the above argument as $i \to \infty$.
Then, we can show that there exists a pair of mappings $(\varphi_n,\psi_n)$ satisfying
\begin{equation}
\limsup_{n \to \infty} D_f(X^n||\psi_n(\varphi_n(X^n))) \le \Delta + \nu,
\end{equation}
and
\begin{IEEEeqnarray}{rCl} \label{eq:t2}
\limsup_{n \to \infty}\frac{1}{n} \log M_n  \le  \limsup_{n \to \infty}\frac{1}{n}H_0(1 - f^{-1}(\Delta +\nu)|X^n).
\end{IEEEeqnarray}
Here, also from the diagonal line argument with respect to $\nu$, we obtain
\begin{IEEEeqnarray}{rCl} 
\limsup_{n \to \infty}\frac{1}{n} \log M_n   \le \lim_{\nu \downarrow 0} \limsup_{n \to \infty}\frac{1}{n}H_0(1 - f^{-1}(\Delta+\nu)|X^n).
\end{IEEEeqnarray}
This completes the proof of the direct part.

(Converse Part:)
We fixed $\nu >0$ arbitrarily.
From Lemma \ref{lemma:converse2},  for any mapping $(\varphi_n,\psi_n)$ satisfying
\begin{equation}
D_f(X^n||\psi_n(\varphi_n(X^n))) \le \Delta + \nu,
\end{equation}
it holds that
\begin{equation}
\frac{1}{n} \log M_n \ge \frac{1}{n}H_0(1 - f^{-1}(\Delta+\nu)|X^n).
\end{equation}
Consequently, we have
\begin{equation}
\limsup_{n \to \infty}D_f(X^n||\psi_n(\varphi_n((X^n))) \le \Delta + \nu
\end{equation}
and
\begin{equation}
\limsup_{n \to \infty}\frac{1}{n} \log M_n \ge \limsup_{n \to \infty} \frac{1}{n}H_0(1 - f^{-1}(\Delta+\nu)|X^n).
\end{equation}
We also use the diagonal line argument \cite{Han}. 
We repeat the above argument as $i \to \infty$ for a sequence $\{\nu_i\}_{i=1}^\infty$ such that $\nu_1 > \nu_2 > \dots > 0$.
Then, for any $(\varphi_n,\psi_n)$ satifying
\begin{equation}
\limsup_{n \to \infty} D_f(X^n||\psi_n(\varphi_n(X^n))) \le \Delta,
\end{equation}
it holds that
\begin{IEEEeqnarray}{rCl}
\limsup_{n \to \infty}\frac{1}{n} \log M_n
&  \ge & \lim_{\nu \downarrow 0}\limsup_{n \to \infty} \frac{1}{n}H_0(1 - f^{-1}(\Delta+\nu)|X^n).
\end{IEEEeqnarray}
This completes the proof of the converse part.

%









\end{document}